\documentclass[3p,twocolumn]{elsarticle}
\usepackage[version=3]{mhchem}
\usepackage{hyperref}
\usepackage{lineno}
\usepackage{textcomp}
\usepackage{amssymb}

\journal{Nuclear Inst. and Methods in Physics Research, B}

\begin{document}

\begin{frontmatter}

\title{Transuranium isotopes at ISAC/TRIUMF\tnoteref{mytitlenote}}
\tnotetext[mytitlenote]{ISAC is the Isotope Separation and ACceleration facility at TRIUMF}

\author[addresslabel1,addresslabel2]{Peter Kunz\corref{cor1}}
\ead{pkunz@triumf.ca}
\cortext[cor1]{Corresponding author. Tel.: +1 (604) 222-7690; Fax: +1 (604) 224-0478}

\author[addresslabel1,addresslabel3]{Jens Lassen}
\author[addresslabel2]{Corina Andreoiu}
\author[addresslabel2,addresslabel4]{Fatima H. Garcia}

\address[addresslabel1]{TRIUMF, 4004 Wesbrook Mall, Vancouver, British Columbia V6T 2A3, Canada}
\address[addresslabel2]{Department of Chemistry, Simon Fraser University, Burnaby, BC V5A 1S6, Canada}
\address[addresslabel3]{Department of Physics, Simon Fraser University, Burnaby, BC V5A 1S6, Canada}
\address[addresslabel4]{Lawrence Berkeley National Laboratory, Nuclear Physics Division, Berkeley, CA, USA}

\begin{abstract}
The production of transuranium isotopes has been demonstrated at the Isotope Separation and ACceleration (ISAC) facility. In particular a laser-ionized $^{239}$Pu beam, extracted from a uranium carbide target, was investigated.
The experimental work was complemented by G\textsc{eant}4 simulations, modelling the impact of secondary particles on the creation of transuranium isotopes through inelastic nuclear reactions. Theoretical production cross sections were derived and compared to experimental results, leading to a discussion on boundary conditions for the release of neptunium and plutonium from ISAC uranium carbide targets.

\end{abstract}

\begin{keyword}
radioactive ion beams \sep
rare isotopes \sep
transuranium elements \sep
plutonium \sep
G\textsc{eant}4 simulations \sep
resonant laser ionization \sep
alpha spectrometry \sep
ISOL technique
\end{keyword}

\end{frontmatter}


\section{Introduction}

The Isotope Separation and ACceleration (ISAC) facility at TRIUMF \cite{dilling_isac_2013} has the capability to produce a wide range of rare isotope beams for research in the fields of nuclear astrophysics, nuclear structure, material science and life science applications. Rare, radioactive isotopes are generated through irradiating suitable target materials with up to 100~$\mu$A of protons from the TRIUMF H$^{-}$ cyclotron. The cyclotron is capable of delivering protons up to 520~MeV though ISAC targets are typically irradiated at an energy of 480~MeV.

The most common proton-induced nuclear reactions involved in the creation of new nuclei fall into the category of inelastic scattering. 
They are classified as fragmentation, spallation, or fission. The latter applies only if the target contains a fissile material. 
All reaction products from these processes are lighter, highly excited fragments of the original target nuclei.
The subsequent de-excitation sheds excess impact energy through the emission of radiation or particles and leads to the formation of rare, mostly radioactive isotopes.
In particular the emission of neutrons during the de-excitation process has the result that the most prominent isotope production channels of spallation and fragmentation reactions are on the neutron-deficient side with regard to the neutron-to-proton ratio of the target element. 
Due to the high neutron-to-proton ratio of fissile elements such as uranium or thorium, fission products are predominantly located on the neutron-rich side of stable isotopes even though neutron-evaporation is part of the fission/de-excitation process.
Apart from neutrons, a plethora of other secondary particles including pions and light, high-energy baryonic fragments are generated as well. Their interactions with target nuclei contribute to the creation of non-negligible quantities of rare isotopes and open production pathways not accessible through proton inelastic scattering. 
Those secondary reactions enable the formation of elements heavier than the elements constituting the target material.
However, in contrast to fusion reactions with high-energy heavy-ion beams for making superheavy elements \cite{armbruster_production_2000}, the impact of secondary reactions in ISAC targets is rather limited. In this paper, we take a closer look on secondary isotope production channels with a particular focus on $^{239}$Pu from uranium targets, putting experimental observations in context with target simulations based on the nuclear transport code G\textsc{eant}4 \cite{allison_recent_2016}.

We will describe the experiment of producing a $^{239}$Pu ion beam from a uranium carbide target through resonant laser ionization, collecting a sample at the ISAC Implantation Station and quantifying the amount of $^{239}$Pu with $\alpha$-spectroscopy.
G\textsc{eant}4 simulations using a hadronic physics model that was specifically adapted to ISAC targets \cite{garcia_calculation_2017} provide information on individual isotope and secondary particle production channels including total reaction cross sections and kinetic energy distributions. 
The combination of experimental and computational results lead to conclusions regarding the most likely production pathway for $^{239}$Pu and the release of neptunium and plutonium from an ISAC uranium carbide target under typical operating conditions.

\section{Experimental}
\label{exp}

The ISAC facility has two target stations which can be operated alternately. While a target in one station is online and bombarded with a proton beam of up to 100~$\rm\mu$A, equivalent to a beam power of up to 48~kW, the next target is prepared for the other station.
The main components of ISAC targets are a 200~mm long cylinder with an inner diameter of 19~mm and an ionizer unit, both entirely made of tantalum.
Ions are extracted in a high-vacuum environment from the ionizer and accelerated up to 60~keV, mass-separated and delivered through an electrostatic beamline network to various experimental stations \cite{bricault_rare_2013}.

A significant fraction of the beam power, typically in the range of several kW, is deposited in the target. Achieving the optimal operating temperature while minimizing temperature gradients within the target container requires an equilibrium between beam heating and external electrical target heating.
In general, high temperatures are beneficial to volatilization and fast release of reaction products, though the maximum temperature is limited by the vapor pressure of target materials and the high-vacuum requirements for efficient ion beam extraction.
The uranium carbide target in this experiment was made of 250 composite ceramic UC$_2$/C discs \cite{kunz_composite_2013} with a total thickness of 0.043~mol/cm$^2$. Typical operating conditions at a calculated temperature of ~1950~\textdegree C were achieved by applying a proton beam current of 6-10~$\rm\mu$A at 480~MeV.
For this experiment, a $^{239}$Pu$^+$ beam was extracted using element-selective resonant laser ionization. This process is discussed in detail in section \ref{rims}. Section \ref{iis} is dedicated to the target irradiation process and the sample collection by implanting the $^{239}$Pu$^+$ beam in an aluminum disc at the ISAC implantation station. The analysis and quantification of the sample is described in section \ref{alpha}.

\subsection{Resonant laser ionization of plutonium}
\label{rims}

\begin{figure}
	\resizebox{0.8\columnwidth}{!}{%
		\includegraphics{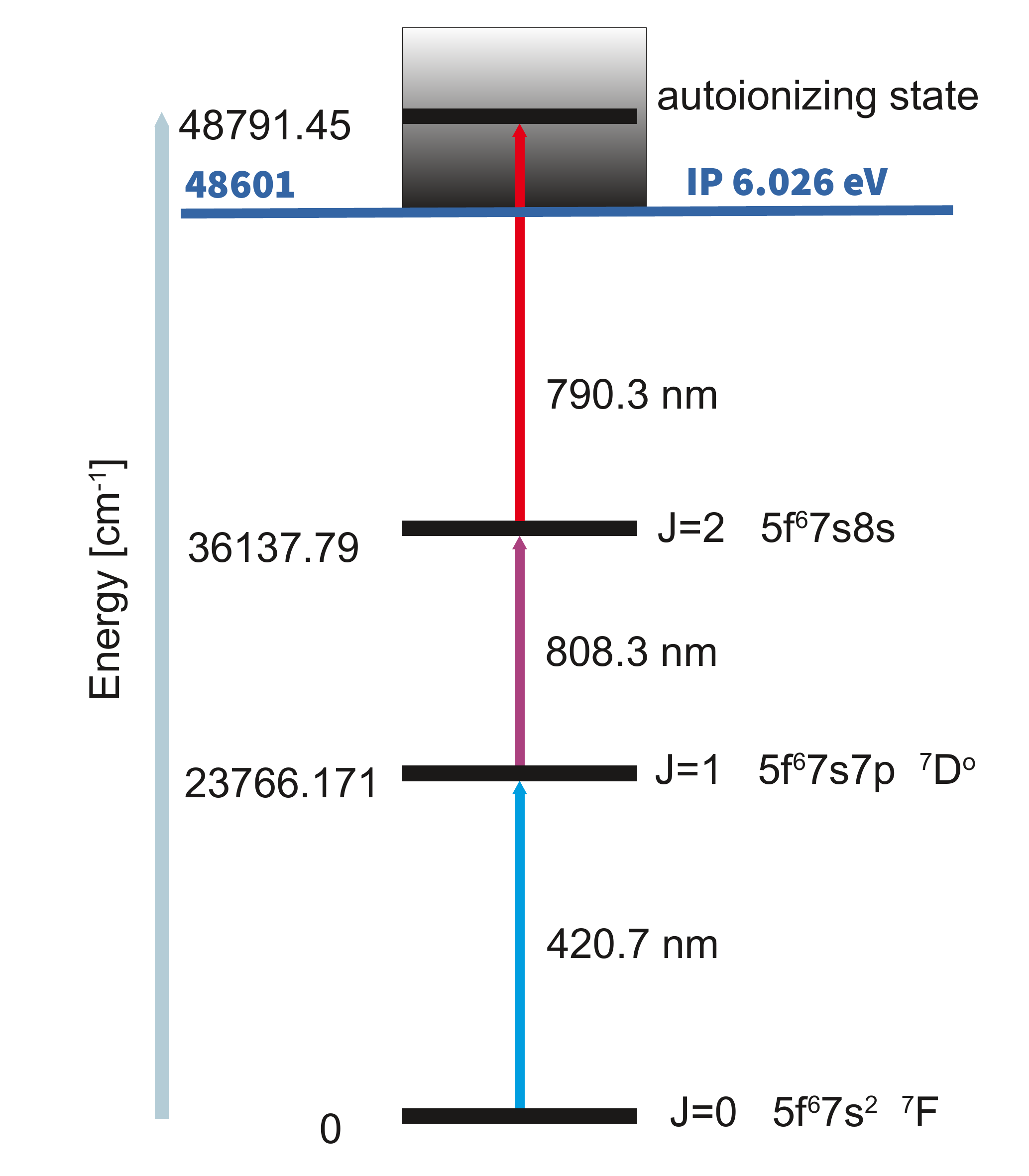}
	}
	\caption{3-step, resonant laser ionization scheme for plutonium. All steps fit well within the operating range of titanium:sapphire lasers.}
	\label{fig:pu_al}       
\end{figure}

Resonance ionization spectroscopy (RIS) pioneered by Hurst and Letokhov \cite{hurst_resonance_1994} was soon adapted for applications in mass spectrometry \cite{zimmermann_resonance_2021} and rare ion beam production at ISOL (Isotope Separation OnLine) facilities \cite{al-khalili_isotope_2006} like ISAC.
Increased ionization efficiency and high element selectivity due to element-specific, multi-step resonant laser excitation of atoms are the main advantages.
ISAC targets are typically equipped with a tantalum-rhenium surface ion source which is operated at a temperature of 2200-2400 \textdegree C \cite{bricault_rare_2013}. In this temperature range alkali elements achieve ionization efficiencies close to 100\% as described by the Saha-Langmuir equation. 
Surface ionization efficiency drops significantly for elements with ionization energies (IPs) above 5~eV. Elements with IP up to 9~eV, including all actinides, are candidates for multi-step, resonant laser ionization using wavelengths from UV to infrared.
Ionization occurs by lifting the valence electron of an atom by step-wise, coherent laser excitation through atomic quantum states.
The final state, above or close the ionization limit, can be a continuum state (non-resonant ionization step), an auto-ionizing state or a Rydberg state. Each option can provide a significant enhancement of ionization efficiency compared to surface ionization, though resonant ionization steps usually require lower laser intensities to saturate the transition with regard to non-resonant ionization.
Plutonium has been an element of interest for RIS as soon as the concept was developed \cite{donohue_detection_1983} and has been a topic of research ever since \cite{galindo-uribarri_high_2021}. A variety of excitation schemes for different laser systems have been investigated yet many remained unpublished due to nuclear non-proliferation concerns. 
The scheme used for this experiment (Fig. \ref{fig:pu_al}) was adopted from Ref. \cite{kunz_efficient_2004}. It is optimized for titanium:sapphire lasers and was easily implemented with the TRIUMF Laser Ion Source (TRILIS) \cite{lassen_current_2017}. 

The laser system used in this work consists of three tunable titanium:sapphire lasers, pumped by a
pulsed, frequency doubled Nd:YAG laser at a repetition rate of 10~kHz with up to 50~W output power and approx. 100~ns pulse length. Each titanium:sapphire laser is pumped with about 10~W power. 
The typical pulse duration of the
titanium:sapphire lasers is 50~ns. All transitions of the excitation scheme shown in Fig.~\ref{fig:pu_al}, including the fundamental of the frequency-doubled 420.7~nm first excitation step, are relatively close to the peak output power within the tuning range of titanium:sapphire lasers, therefore, providing sufficient intensity to saturate all transitions.
The three laser beams were spatially and temporally overlapped in the standard 3~mm inner diameter tantalum-rhenium ionizer tube of the ISAC target. The laser bandwidth of 4-6~GHz is sufficient to cover the Doppler-broadening of transitions in the hot ionizer tube which, with regard to plutonium, functions like a neutral atomic vapor source.

Using this configuration with optimized wavelengths of 841.526~nm (frequency doubled) at 170~mW for the first step, 808.305~nm (600~mW) for the second step and 790.287~nm (2~W) for the final step to an autoionizing state, a rate of ~2.4$\cdot 10^{7}$~ions/sec was detected on a channeltron in the ISAC beamline, a count rate ~36 times higher than the non-resonant rate with lasers turned off.
This is consistent with a theoretical surface ionization efficiency of plutonium in the order of 1\% calculated with the Saha-Langmuir equation \cite{langmuir_thermionic_1925} for a rhenium surface ion source at 2000\textdegree~C, considering uncertainties related to the exact knowledge of the ionizer temperature and differences in ion extraction efficiency between surface and laser ionization.
It cannot be ruled out that part of the non-resonant beam consists of $^{239}$Np$^+$ or isobaric molecules, for example $^{220}$Ra$^{19}$F$^+$ and $^{220}$Ac$^{19}$F$^+$, with relatively low ionization energies. 
The available beamtime did not allow a control measurement without lasers. Therefore, the observed resonance enhancement factor can only be viewed as a lower limit.

\subsection[$^{239}$Pu$^+$ implantation]{$^{239}$Pu$^+$ implantation}
\label{iis}

\begin{figure}
	\resizebox{1.0\columnwidth}{!}{%
		\includegraphics{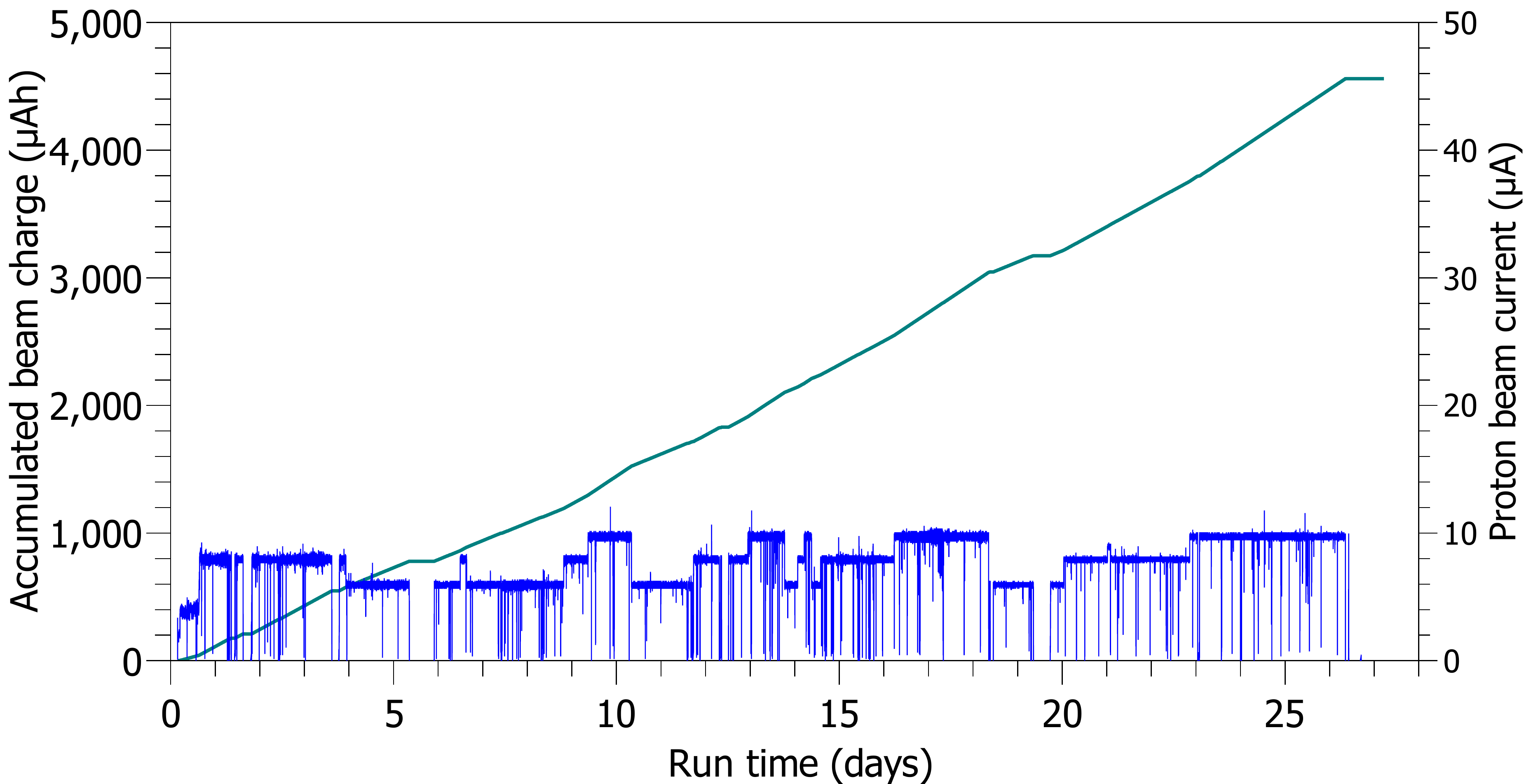}
	}
	\caption{Proton beam irradiation of uranium carbide target ITW-TM1-UC6 over 28 days. The beam current is depicted by the blue line and the green line shows the accumulated beam charge.}
	\label{fig:uc6}       
\end{figure}

Preceding the extraction of a plutonium beam from uranium carbide target \#6 (ITW-TM1-UC6) and its collection at the ISAC Implantation Station (IIS), the target was irradiated with typically 6-10 $\mu$A of 480~MeV protons over a period of 28 days. A variety of rare ion beams were extracted during the run period for various experiments. The target was kept at an average temperature of 1920~\textdegree C. 
This value is associated with the core temperature where the proton beam traverses through the target. It is calculated by a heat transfer model that takes external electrical heating, proton beam power deposition, thermal conductivity of the target material and the emissivity of the target container into account \cite{dombsky_isac_2003}.
As shown in Fig.~\ref{fig:uc6}, at the end of the irradiation period a total proton beam charge of 4560 $\mu$Ah had been accumulated on the target. The median proton beam current was 8~$\mu$A.

The ISAC Implantation Station is a branch of the ISAC electro-static beamline network a few meters downstream of the ISAC high-resolution mass separator. It is dedicated to the collection of long-lived radioactive isotopes for a variety of applications such as precision half-life measurements or quantum detector developments that require ion beam implantations at specific intensities and energies \cite{friedrich_limits_2021}. A major application area is the collection of isotopes for nuclear medicine research \cite{kunz_medical_2020}\cite{fiaccabrino2021}. 
The IIS provides the infrastructure for ion beam positioning, rastering and focusing on a custom implantation target that can be attached to a vacuum port within a user-designed vacuum chamber. Options for biasing and beam diagnostics are available as well. Examples for user-defined configurations are given in the references above.

For this experiment, the plutonium ion beam was extracted at the end of the irradiation period after the proton beam was turned off. Keeping the external target heating unchanged, the target temperature was according to the heat transfer model \cite{dombsky_isac_2003} approximately 200~\textdegree C lower than the nominal operating temperature due to the lack of power input from the proton beam.
After establishing resonant laser ionization of plutonium as described in section \ref{rims}, a 20~keV ion beam was directed to the IIS where it was implanted for 12h~14min (44040~s) on a aluminium disc with a diameter of 12.7~mm in a simple vacuum chamber attached to the IIS user port without further diagnostics.
During the implantation period the ion beam current was frequently monitored with a Faraday cup approximately 1~m upstream of the implantation target. The cup was operated in reverse-bias mode and could therefore not provide a calibrated current output but verified that the ion beam intensity dropped to 62\% of the maximum beam intensity during the implantation period. The available data does not allow a conclusion whether the drop is the result of depletion of the target inventory, changes in the ion beam transmission or the ionization efficiency. The result based on the analysis of accumulated activity, discussed in section \ref{alpha}, is therefore an average over the complete implantation period.

\subsection[alpha-Spectroscopy of $^{239}$Pu]{$\alpha$-Spectroscopy of $^{239}$Pu}
\label{alpha}

\begin{figure}
	\resizebox{1.0\columnwidth}{!}{%
		\includegraphics{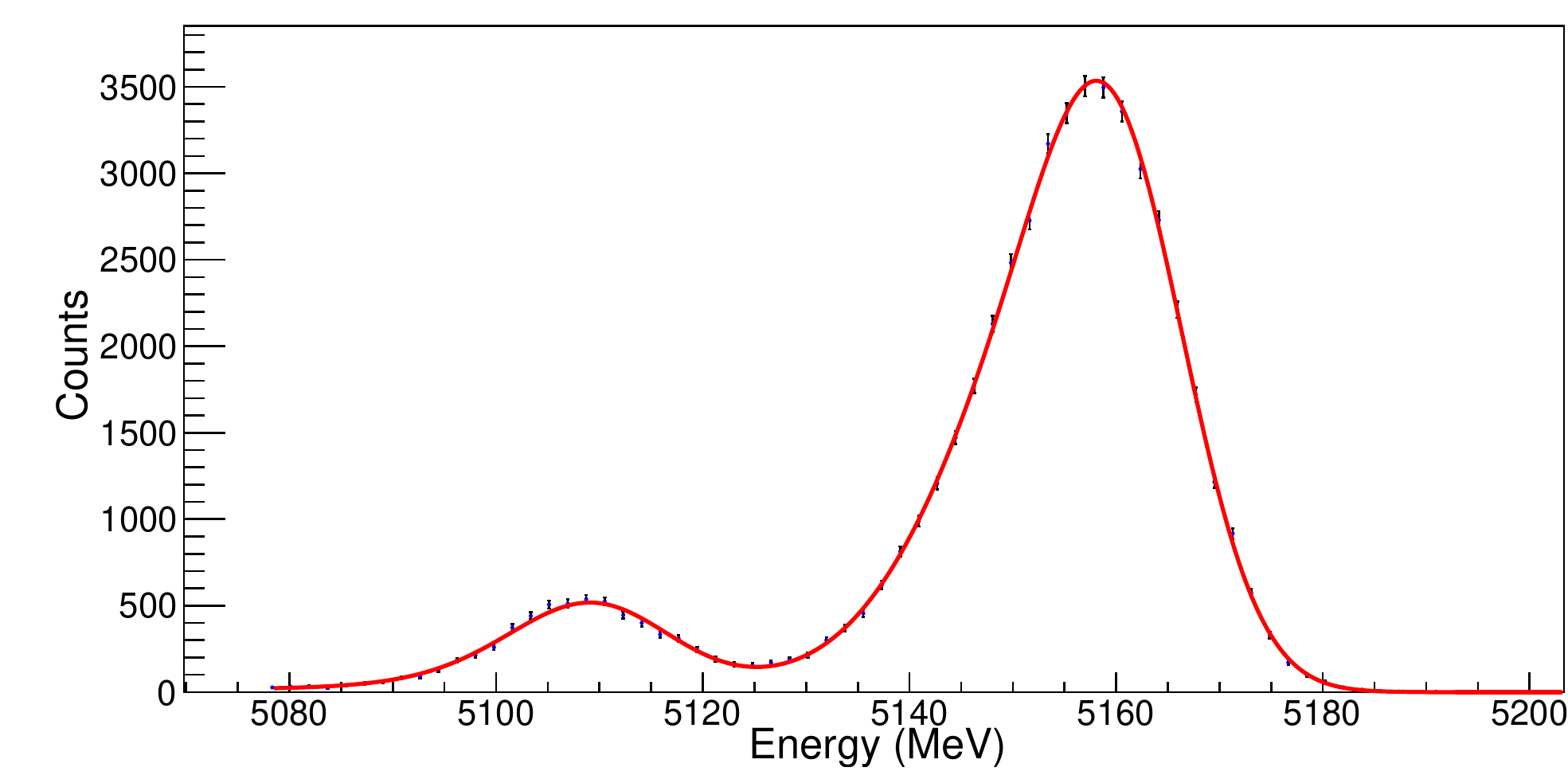}
	}
	\caption{High-resolution alpha decay showing the 3 major alpha lines of $^{239}$Pu. The two strongest lines at 5156.59~keV (70.77\%) and 5144.3~keV (17.11\%) are not resolved. The weaker line at 5105.5~keV (11.94\%) is the smaller peak on the left. The red line is a fit to the data points. The decay events were collected over a period of 20 days.}
	\label{fig:pu239_alpha}       
\end{figure}

$^{239}$Pu decays 100\% through alpha emission (a very small spontaneous fission decay branch is negligible for this investigation). 99.82\% of the alpha decays are part of 3 major channels at 5156.59~keV (70.77\%), 5144.3~keV (17.11\%) and 5105.5~keV (11.94\%), respectively.
The isotope has a half-life of 24114~years. An activity of 1~Bq is equivalent to 1.098$\cdot 10^{12}$ atoms. The observation of the channeltron count rate (\ref{rims}) and also the not-calibrated Faraday cup output (\ref{iis}) indicate that an accumulated activity over the given implantation period should be expected to be in the same order.
After the implantation, the target was retrieved and the deposited isotopes were investigated with using a ORTEC Soloist alpha spectrometer with a low-background detector (BU-024-600AS).
According to a SRIM \cite{ziegler_srim_2010} simulation the vast majority of plutonium ions are stopped in the aluminium sample disc at a depth around 15.3~nm. 
Taking the straggling effect into account, almost none are implanted deeper than 30~nm.
Alpha particles emitted from Pu atoms embedded at this depth experience an average energy loss of $\sim$2.5~keV which is almost an order of magnitude smaller than the warranted detector resolution of 24~keV and therefore not relevant in this context.
Several measurements were performed. A measurement with the sample close to the detector, resulting in a high detection efficiency (28(1)\%) but lower resolution was made over a period of 24 hours to determine the total activity implanted on the target. In another measurement, decay data was collected for 20~days with a larger distance between source and detector. At the expense of reduced efficiency the smaller solid angle between source and detector improved the resolution of the alpha decay spectrum shown in Fig.~\ref{fig:pu239_alpha}. Besides the two peaks depicted in the figure, no other significant alpha lines were observed.
The larger peak consists of the two strongest lines at 5156.59~keV (70.77\%) and 5144.3~keV (17.11\%) wich could not be resolved. The smaller peak is third decay at 5105.5~keV (11.94\%). In order to establish a $^{239}$Pu identification independent of the response to resonant laser ionization an analytical function for fitting peaks in alpha-particle spectra from Si detectors \cite{bortels_analytical_1987} was minimized to the data set in Fig.~\ref{fig:pu239_alpha} using the MINUIT algorithm \cite{james_minuit_1975}. The fit result at a reduced $\chi^2$ of 0.996 is in good agreement with the literature values above at 10.7(2)\% intensity for the 5105.5~keV line and a combined intensity of 87.4(1.2)\% for the two merged lines.

The 24h high-efficiency measurement revealed a total activity based on the counts in the alpha peaks discussed above of 0.758(24)~Bq which corresponds to an average yield over the implantation period of 1.89(6)$\cdot 10^{7}$ $^{239}Pu^{+}$ ions/second. The error associated with the result is based on the Gaussian propagation of statistical errors and the uncertainty of the calibration source.

\section{Target simulations}
\label{sim}


The vast majority of nuclides produced in ISAC uranium carbide targets are the result of spallation, fragmentation and fission.
$^{239}$Pu having a higher baryon number than its progenitor $^{238}$U obviously must be created through different channels involving secondary particles stemming from proton-induced nuclear reactions.
G\textsc{eant}4 simulations of an ISAC uranium carbide target were conducted to study the production of isotopes above mass number 238 through secondary reactions.
The simulation data discussed in this section was obtained using a standardized model of ISAC targets \cite{garcia_calculation_2017} that was developed to guide ISAC facility users in planning their experiments. The results in the form of normalized isotope production rates  for a variety of target elements in their natural isotope composition at the typical proton beam energy of 480~MeV are publicly available in the TRIUMF Isotope Database \cite{isacyielddb}.
The database does not contain simulation results of composite materials such as carbides or oxides. The vast majority of production rates can be extrapolated by combining and scaling the rates from individual elements according to their stoichiometric ratio in the target material. For this work, the carbon content of a composite ceramic uranium carbide target \cite{kunz_composite_2013} was included to obtain a more realistic picture of secondary light particle production. Composite ceramic target discs contain some excess carbon in their carbide fraction, though most of the carbon stems from the graphite backing foils which are crucial for structural stability and improved thermal conductivity.
The combined inventory amounts to a C:U stoichiometric ratio of 8:1.
\begin{figure}
	\resizebox{1.0\columnwidth}{!}{%
		\includegraphics{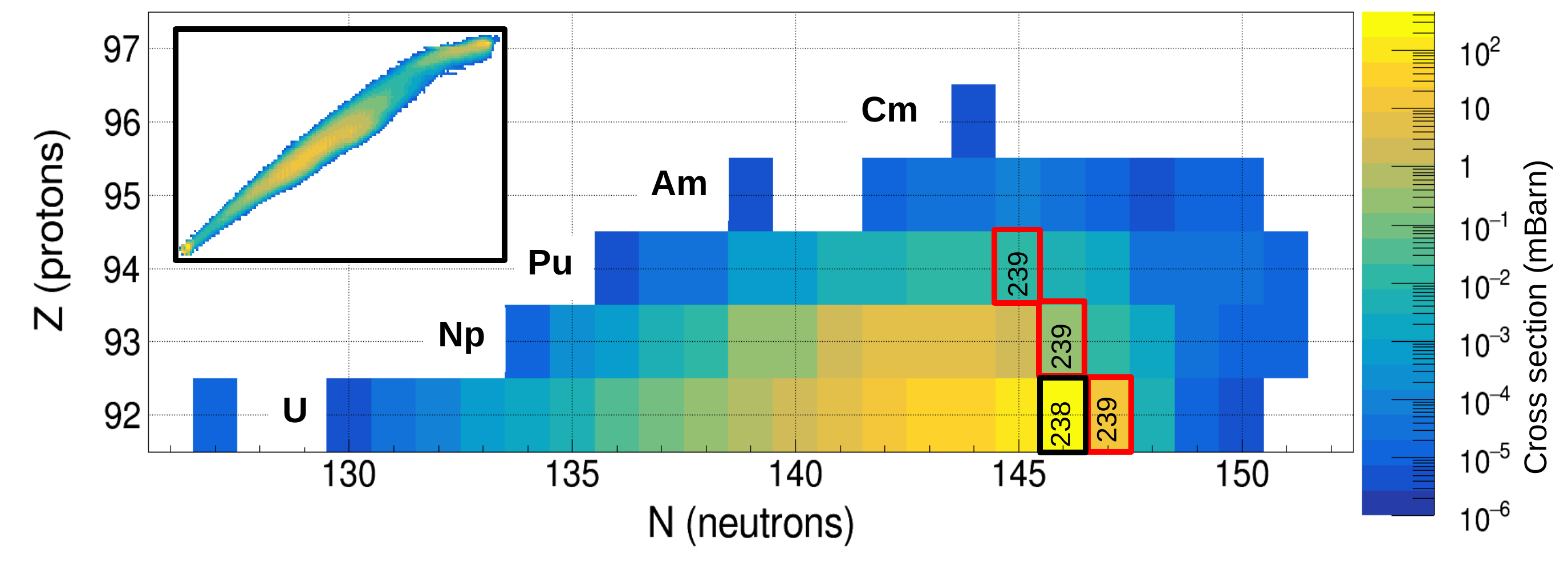}
	}
	\caption{G\textsc{eant}4 simulation of production cross sections for transuranium isotopes from a standardized ISAC uranium carbide target. The chart shows uranium and transuranium direct reaction products of $^{238}$U (black sqare). The indirect production channel of $^{239}$Pu (see Eq.~\ref{eq:decay_u239}) is highlighted (red squares).
	This area is the upper tip of the full range of isotope production from UC$_8$ (see insert).}
	\label{fig:g4_uc8}       
\end{figure}

The simulation models the target in a simplified geometry as a cylinder with a diameter of 19~mm and a length of 500~mm. The standard target thickness of 0.05~mol/cm$^2$ is set by filling the cylinder with target material of matching density. 
These parameters, dimensions as well as target thickness, are fairly similar to target ITW-TM1-UC6 specified in section \ref{exp}.
For this work the model was refined to extract relevant process data such as kinetic energies of primary and secondary particles, time stamps and reaction types.
Using G\textsc{eant} version 11.00 with the Liege Intra-Nuclear Cascade Model \cite{rodriguez_2017} implemented with physics list QGSP\_INCLXX\_HP and combined with the ABLA evaporation code \cite{kelic_abla07_2008} a data set tracking $10^{10}$ 480~MeV protons impinging on the target along the cylindrical axis (7~mm FWHM) was obtained.
Figure \ref{fig:g4_uc8} provides an overview of isotope production cross sections extracted from the data set in the region of interest and highlights the most relevant isotopes for this work.

\begin{figure}
	\resizebox{1.0\columnwidth}{!}{%
		\includegraphics{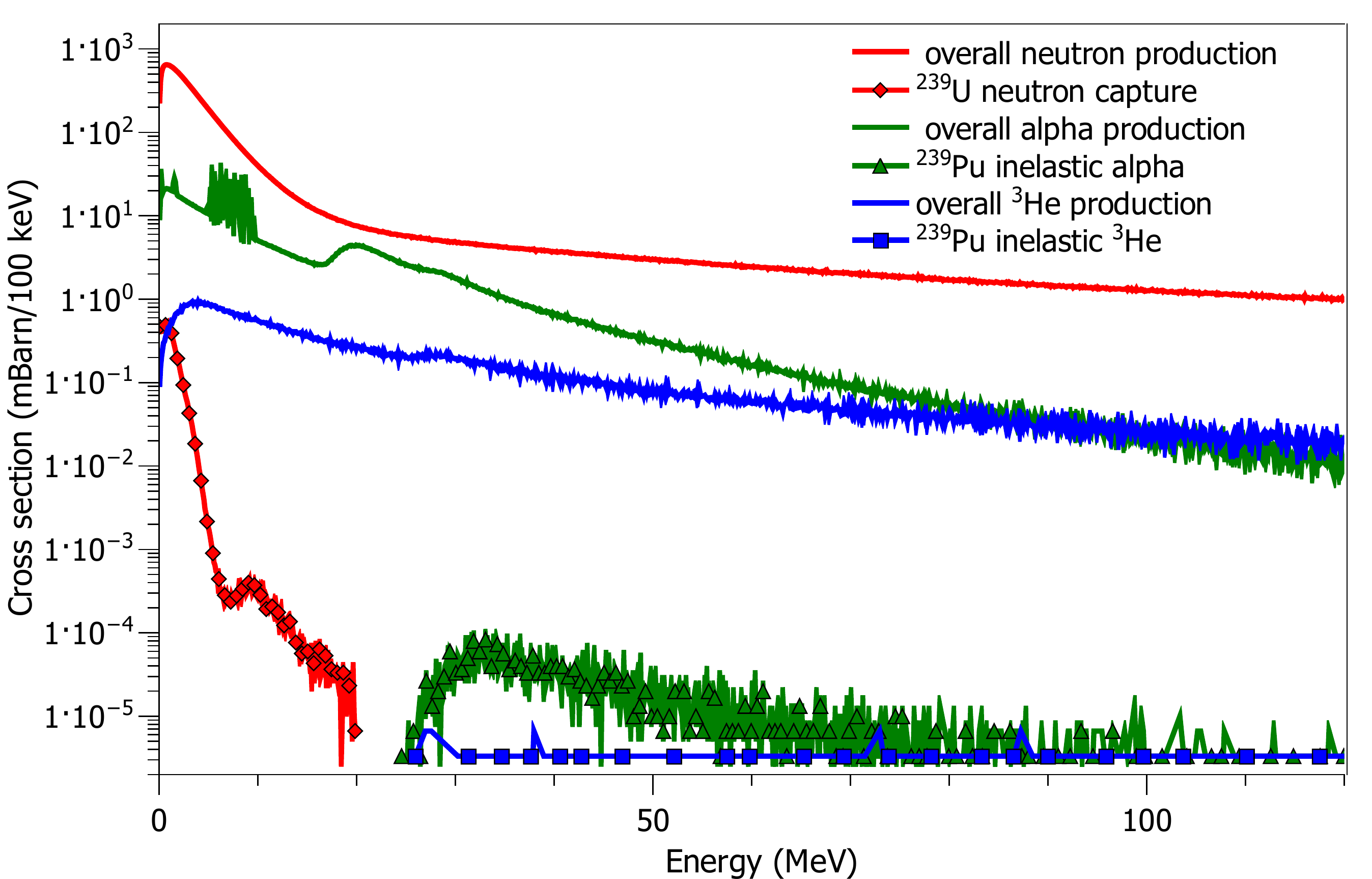}
	}
	\caption{G\textsc{eant}4 simulation results of energy dependent total cross sections for $^{239}$Pu production from a standardized ISAC uranium carbide target. Direct production channels: alpha inelastic scattering (green $\triangle$), overall alpha production (solid green line); $^{3}$He inelastic scattering (blue $\square$), overall $^{3}$He production (solid blue line). Indirect production channel: neutron capture $^{239}$U (red $\Diamond$), overall neutron production (solid red line).}
	\label{fig:pu239_cross}       
\end{figure}

There are two ways to generate $^{239}$Pu from a uranium carbide target, either directly through inelastic reactions with secondary particles such as $^{3,4}$He or indirectly through $^{239}$Np decay created from p,$^{2,3}$H inelastic reactions and the decay of $^{239}$U (Eq. \ref{eq:decay_u239}) which is produced via neutron capture. 
\begin{equation}
\label{eq:decay_u239}
\begin{split}
\ce{^{239}U ->[{\beta^{-}}][{23.45 m}] ^{239}Np ->[{\beta^{-}}][{2.356 d}] ^{239}Pu }
\end{split}
\end{equation}
Figure \ref{fig:pu239_cross} summarizes the simulation results related to neutrons, $^{3}$He and $^{4}$He (alpha particles). It depicts the overall total production cross sections for this specific target configuration as a function of energy. Furthermore, it compares these results with the cross sections for $^{239}$Pu production involving the same particles.

The energy spectrum of neutrons resulting in the creation of $^{239}$U as a precursor for $^{239}$Pu aligns well with the distribution of all neutrons in the target. The ratio of the sum of neutrons producing $^{239}$U over the sum of all neutrons is  $2.89\cdot 10^{-4}$.

%

In contrast, the inelastic reaction $^{238}$U($^{4}$He,3n)$^{239}$Pu utilizes a significant smaller fraction ($4.73\cdot 10^{-6}$) of alpha particles with kinetic energies higher than $\sim$25~MeV. 
The same observation can be made for the reaction $^{238}$U($^{3}$He,2n)$^{239}$Pu, though the $^{239}$Pu production cross section is almost a factor of 20 smaller than for alpha inelastic scattering.
The total cross sections $\sigma$ derived from the G\textsc{eant}4 simulation are listed in table \ref{tab:cross_sections}.
Also contributing to the indirect production channel is $^{239}$Np from inelastic proton, deuterium and tritium scattering. The combined cross section of $1.81\cdot 10^{-1}$~mBarn of these reactions is small compared to neutron capture and therefore not considered in this context.

\begin{table}[h!]
	\begin{center}
		\caption{Total ISAC uranium carbide target cross sections $\sigma$ for overall neutron, alpha and $^{3}$He production and direct as well as indirect creation of $^{239}$Pu.}
		\label{tab:cross_sections}
		\begin{tabular}{c|c|c} 
			$\sigma$ (mBarn)&overall &$^{239}$Pu\\
			\hline
			neutron & $3.13\cdot 10^{4}$ &  $9.04\cdot 10^{0}$\\
			alpha & $2.38\cdot 10^{3}$ &  $1.13\cdot 10^{-2}$\\
			$^{3}$He& $1.92\cdot 10^{2}$ &  $6.18\cdot 10^{-4}$\\
		\end{tabular}
	\end{center}
\end{table}

The production rate of $^{239}$Pu in the actual target is given by Eq. \ref{eq:prod}, where
$\Phi_{p+}$ is the proton flux (protons/s) of energy $E$, $\sigma$ the total production cross section (mBarn) at the same energy and $N_{U}$ the thickness of uranium (atoms/cm$^2$).

\begin{equation}
	\label{eq:prod}
	P(^{239}Pu) = \Phi_{p+}(E)\; \sigma(E)\;  N_{U}
\end{equation}

Using the calculated production cross sections and the ITW-TM1-UC6 target parameters (8~$\mu$A p+; 0.043~mol/cm$^2$) the production rates for $^{239}$Pu through neutron capture and alpha inelastic scattering are $1.1\cdot 10^{10}$/s and $1.4\cdot 10^{7}$/s, respectively. The $^{3}$He channel does not contribute significantly.
The interpretation of these results with respect to the experimental data will be discussed in the following section.

\section{Results and Discussion}
\label{results}

The yield of specific ion beams from a target depends on the in-target production rate $P$ (Eq. \ref{eq:prod}) multiplied by efficiency factors for release $\epsilon_R$, ionization $\epsilon_I$ and beam transportation $\epsilon_T$.
\begin{equation}
\label{eq:yield}
Yield = P\; \epsilon_R\; \epsilon_I\; \epsilon_T
\end{equation}
Transport and ionization are usually constant factors whereas the release efficiency is governed by physical target properties (dimensions, thickness, porosity) and chemical reactions between released isotope and target material.
The time constants for potential chemical reactions and migration of isotopes to the ion source are strongly dependent on the temperature distribution within the target. 
They are associated with an average release time which can have a great impact on the release efficiency if it is longer or on the same scale of the half-life of the released isotope.
A relatively long release time can lead to build-up of in-target inventory and generation of secondary isotopes through radioactive decay.

The potential build-up of a $^{239}$Pu or $^{239}$Np inventory can be described by a modified Bateman equation.
\begin{equation}
\label{eq:bateman}
\frac{dN_i}{dt} = \lambda_{i-1}  N_{i-1} -  \lambda_{i}  N_{i} + P_i - R_i 
\end{equation}
The formation rate of isotope $N_i$ is determined by the decay constants of itself $\lambda_i$ and its precursor $\lambda_{i-1}$. 
In addition it depends on a constant production rate $P_i$ (see Eq. \ref{eq:prod} )and a release rate $R_i$ that takes account of isotopes escaping through the ionizer.
The experimental data outlined in section \ref{exp} is not sufficient to pin down release rates in relation to production and decay.
However, some boundary conditions can be explored by combining the measured $^{239}$Pu yield with simulation results.
First, it is reasonable to expect that $^{239}$U, as a chemical equivalent to natural uranium isotopes, is not released. Assuming otherwise contradicts the observation that the target material itself was stable over weeks of operation.
Using Eq. \ref{eq:bateman} with the production rates from section \ref{sim} under the assumption that neither neptunium nor plutonium is released ($R_i = 0$) the total inventory of $^{239}$Np and $^{239}$Pu after 28 days of irradiation would be $3.2\cdot 10^{15}$ and $2.3\cdot 10^{16}$ atoms or $1.1\cdot 10^{10}$ and $2.1\cdot 10^{4}$ Bequerel, respectively.
In this theoretical scenario the direct production channel contributes only 0.145\% to the final $^{239}$Pu inventory.
However, an unrelated yield measurement of $^{238}$Np \cite{isacyielddb} and the results reported in section \ref{exp} confirm that both elements are released from uranium carbide targets, though the actual release efficiency in each case is unknown.
In another scenario under which 100\% of plutonium is promptly released, direct production can be ruled out since the yield of $\sim 2\cdot 10^{7}$ ions/s was measured after the proton beam was turned off. It is slightly higher than the calculated direct production rate and for that reason cannot account for the measured yield.

\begin{figure}
	\resizebox{1.0\columnwidth}{!}{%
		\includegraphics{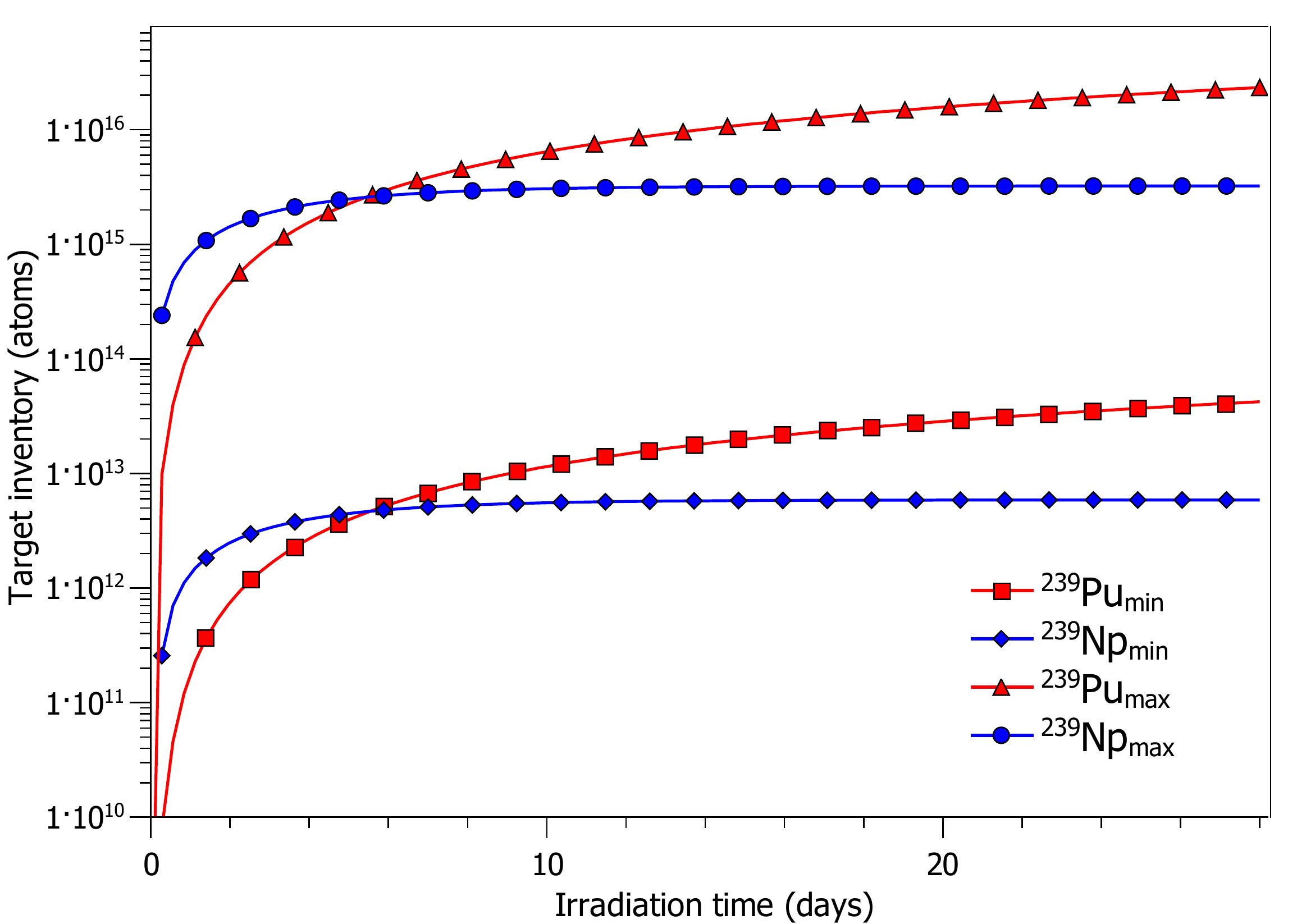}
	}
	\caption{In-target build-up of $^{239}$Np and $^{239}$Pu over irradiation period of 28 days based on Eq.~\ref{eq:bateman} and production rates from section \ref{sim}. Results with subscript \textit{max} assume no release from target ($R_i = 0$). Results with subscript \textit{min} assume $^{239}$Np release from target ($R_i = 1.098\cdot 10^{10}$~atoms/s). This condition leaves a $^{239}$Np activity at the end of the period roughly equivalent to the measured $^{239}$Pu yield.}
	\label{fig:inventory}       
\end{figure}

That leaves the indirect production channel as the major source of plutonium, which requires that the intermediate isotope $^{239}$Np is retained in the target to a certain extent so it can act as a $^{239}$Pu generator after the proton irradiation has stopped.
Assuming that $\epsilon_R \epsilon_I \epsilon_T = 1$, the minimum required amount would be the same $^{239}$Np activity as the measured $^{239}$Pu yield which is equivalent to  $\sim 5.9\cdot 10^{12}$ atoms at the end of the run period or equivalent to a $^{239}$Np release rate of $R_i = 1.098\cdot 10^{10}$~atoms/s. That is only 0.18\% of the maximum inventory noted above.
Considering that neither ionization or transport can have an efficiency of 100\%, the release of both, neptunium and plutonium must scale somewhere between the boundary conditions outlined above and illustrated in Fig.~\ref{fig:inventory}.

\section{Summary and Conclusion}
A pure $^{239}$Pu ion beam was extracted from a ISAC uranium carbide target that had been irradiated for over 28 days with 480 MeV protons. It was identified through resonant laser ionization spectroscopy and alpha spectrometry. The beam intensity was determined with direct ion beam and radiometric yield measurements.
A G\textsc{eant}4 simulation was developed to model various channels of $^{239}$Pu creation including the impact of secondary particles. Production cross sections and rates were derived from the simulation data sets.
As discussed in the previous section, release efficiencies of neptunium and plutonium could not be exactly determined. It will require further investigations to understand the impact of chemical and thermodynamic processes involved in the release of both elements. 

It has been demonstrated that $^{239}$Pu, the heaviest isotope investigated at the ISAC facility so far, can be produced in sufficient quantities for potential RIB experiments.
The combined conclusions from measurements and simulations indicate that significantly higher yields can be achieved through inventory build-up by irradiating a cold target. 
 
\section*{Acknowledgements}
TRIUMF receives federal funding via a contribution agreement  through  the  National  Research  Council  of  Canada.   We also acknowledge additional support through the New Frontiers in Research Fund - Exploration  NFRFE-2019-00128 and Discovery Grants from  the  Natural  Sciences  and  Engineering  Research  Council  of  Canada, (NSERC):  SAPIN-2021-00030  (P.  Kunz)  and SAPIN/00039/2017 (J. Lassen).  Many thanks to the ISAC operations crew for their support and the TRIUMF Radiation Protection Group for their help managing fissile materials.

\bibliography{pu}

\begin{thebibliography}{10}
\expandafter\ifx\csname url\endcsname\relax
  \def\url#1{\texttt{#1}}\fi
\expandafter\ifx\csname urlprefix\endcsname\relax\def\urlprefix{URL }\fi
\expandafter\ifx\csname href\endcsname\relax
  \def\href#1#2{#2} \def\path#1{#1}\fi

\bibitem{dilling_isac_2013}
J.~Dilling, R.~Kr\"{u}cken, G.~Ball,
  \href{http://link.springer.com/10.1007/s10751-013-0877-7}{{ISAC} overview},
  Hyperfine Interactions 225~(1-3) (2014) 1--8.
\newblock \href {http://dx.doi.org/10.1007/s10751-013-0877-7}
  {\path{doi:10.1007/s10751-013-0877-7}}.
\newline\urlprefix\url{http://link.springer.com/10.1007/s10751-013-0877-7}

\bibitem{armbruster_production_2000}
P.~Armbruster,
  \href{https://www.annualreviews.org/doi/10.1146/annurev.nucl.50.1.411}{On the
  {Production} of {Superheavy} {Elements}}, Annual Review of Nuclear and
  Particle Science 50~(1) (2000) 411--479.
\newblock \href {http://dx.doi.org/10.1146/annurev.nucl.50.1.411}
  {\path{doi:10.1146/annurev.nucl.50.1.411}}.
\newline\urlprefix\url{https://www.annualreviews.org/doi/10.1146/annurev.nucl.50.1.411}

\bibitem{allison_recent_2016}
J.~Allison, K.~Amako, J.~Apostolakis, P.~Arce, M.~Asai, T.~Aso, E.~Bagli,
  A.~Bagulya, S.~Banerjee, G.~Barrand, B.~R. Beck, A.~G. Bogdanov, D.~Brandt,
  J.~M.~C. Brown, H.~Burkhardt, P.~Canal, D.~Cano-Ott, S.~Chauvie, K.~Cho,
  G.~A.~P. Cirrone, G.~Cooperman, M.~A. Cortés-Giraldo, G.~Cosmo, G.~Cuttone,
  G.~Depaola, L.~Desorgher, X.~Dong, A.~Dotti, V.~D. Elvira, G.~Folger,
  Z.~Francis, A.~Galoyan, L.~Garnier, M.~Gayer, K.~L. Genser, V.~M. Grichine,
  S.~Guatelli, P.~Guèye, P.~Gumplinger, A.~S. Howard, I.~Hřivnáčová,
  S.~Hwang, S.~Incerti, A.~Ivanchenko, V.~N. Ivanchenko, F.~W. Jones, S.~Y.
  Jun, P.~Kaitaniemi, N.~Karakatsanis, M.~Karamitros, M.~Kelsey, A.~Kimura,
  T.~Koi, H.~Kurashige, A.~Lechner, S.~B. Lee, F.~Longo, M.~Maire, D.~Mancusi,
  A.~Mantero, E.~Mendoza, B.~Morgan, K.~Murakami, T.~Nikitina, L.~Pandola,
  P.~Paprocki, J.~Perl, I.~Petrović, M.~Pia, W.~Pokorski, J.~M. Quesada,
  M.~Raine, M.~A. Reis, A.~Ribon, A.~Ristić~Fira, F.~Romano, G.~Russo,
  G.~Santin, T.~Sasaki, D.~Sawkey, J.~I. Shin, I.~I. Strakovsky, A.~Taborda,
  S.~Tanaka, B.~Tomé, T.~Toshito, H.~N. Tran, P.~R. Truscott, L.~Urban,
  V.~Uzhinsky, J.~M. Verbeke, M.~Verderi, B.~L. Wendt, H.~Wenzel, D.~H. Wright,
  D.~M. Wright, T.~Yamashita, J.~Yarba, H.~Yoshida,
  \href{https://linkinghub.elsevier.com/retrieve/pii/S0168900216306957}{Recent
  developments in {Geant4}}, Nuclear Instruments and Methods in Physics
  Research Section A: Accelerators, Spectrometers, Detectors and Associated
  Equipment 835 (2016) 186--225.
\newblock \href {http://dx.doi.org/10.1016/j.nima.2016.06.125}
  {\path{doi:10.1016/j.nima.2016.06.125}}.
\newline\urlprefix\url{https://linkinghub.elsevier.com/retrieve/pii/S0168900216306957}

\bibitem{garcia_calculation_2017}
F.~H. Garcia, C.~Andreoiu, P.~Kunz,
  \href{https://linkinghub.elsevier.com/retrieve/pii/S0168583X17308893}{Calculation
  of in-target production rates for isotope beam production at {TRIUMF}},
  Nuclear Instruments and Methods in Physics Research Section B: Beam
  Interactions with Materials and Atoms 412 (2017) 174--179.
\newblock \href {http://dx.doi.org/10.1016/j.nimb.2017.09.023}
  {\path{doi:10.1016/j.nimb.2017.09.023}}.
\newline\urlprefix\url{https://linkinghub.elsevier.com/retrieve/pii/S0168583X17308893}

\bibitem{bricault_rare_2013}
P.~G. Bricault, F.~Ames, M.~Dombsky, P.~Kunz, J.~Lassen,
  \href{http://link.springer.com/10.1007/s10751-013-0880-z}{Rare isotope beams
  at {ISAC} - target \& ion source systems}, Hyperfine Interactions 225~(1-3)
  (2014) 25--49.
\newblock \href {http://dx.doi.org/10.1007/s10751-013-0880-z}
  {\path{doi:10.1007/s10751-013-0880-z}}.
\newline\urlprefix\url{http://link.springer.com/10.1007/s10751-013-0880-z}

\bibitem{kunz_composite_2013}
P.~Kunz, P.~Bricault, M.~Dombsky, N.~Erdmann, V.~Hanemaayer, J.~Wong,
  K.~L\"{u}tzenkirchen,
  \href{http://linkinghub.elsevier.com/retrieve/pii/S0022311513006831}{Composite
  uranium carbide targets at {TRIUMF:} development and characterization with
  {SEM}, {XRD}, {XRF} and {L}-edge densitometry}, Journal of Nuclear Materials
  440~(1-3) (2013) 110--116.
\newblock \href {http://dx.doi.org/10.1016/j.jnucmat.2013.04.065}
  {\path{doi:10.1016/j.jnucmat.2013.04.065}}.
\newline\urlprefix\url{http://linkinghub.elsevier.com/retrieve/pii/S0022311513006831}

\bibitem{hurst_resonance_1994}
G.~S. Hurst, V.~S. Letokhov,
  \href{http://physicstoday.scitation.org/doi/10.1063/1.881420}{Resonance
  {Ionization} {Spectroscopy}}, Physics Today 47~(10) (1994) 38--45.
\newblock \href {http://dx.doi.org/10.1063/1.881420}
  {\path{doi:10.1063/1.881420}}.
\newline\urlprefix\url{http://physicstoday.scitation.org/doi/10.1063/1.881420}

\bibitem{zimmermann_resonance_2021}
M.~Savina, R.~Trappitsch,
  \href{https://onlinelibrary.wiley.com/doi/10.1002/9783527682201.ch6}{Resonance
  {Ionization} {Mass} {Spectrometry} {RIMS}: {Fundamentals} and {Applications}
  {Including} {Secondary} {Neutral} {Mass} {Spectrometry}}, in: R.~Zimmermann,
  L.~Hanley (Eds.), Photoionization and {Photo}‐{Induced} {Processes} in
  {Mass} {Spectrometry}, 1st Edition, Wiley, 2021, pp. 215--244.
\newblock \href {http://dx.doi.org/10.1002/9783527682201.ch6}
  {\path{doi:10.1002/9783527682201.ch6}}.
\newline\urlprefix\url{https://onlinelibrary.wiley.com/doi/10.1002/9783527682201.ch6}

\bibitem{al-khalili_isotope_2006}
P.~Van~Duppen, \href{http://link.springer.com/10.1007/3-540-33787-3_2}{Isotope
  {Separation} {On} {Line} and {Post} {Acceleration}}, in: J.~Al-Khalili,
  E.~Roeckl (Eds.), The {Euroschool} {Lectures} on {Physics} with {Exotic}
  {Beams}, {Vol}. {II}, Vol. 700, Springer Berlin Heidelberg, 2006, pp. 37--77,
  series Title: Lecture Notes in Physics.
\newblock \href {http://dx.doi.org/10.1007/3-540-33787-3_2}
  {\path{doi:10.1007/3-540-33787-3_2}}.
\newline\urlprefix\url{http://link.springer.com/10.1007/3-540-33787-3_2}

\bibitem{donohue_detection_1983}
D.~L. Donohue, J.~P. Young,
  \href{https://pubs.acs.org/doi/abs/10.1021/ac00253a046}{Detection of
  plutonium by resonance ionization mass spectrometry}, Analytical Chemistry
  55~(2) (1983) 378--379.
\newblock \href {http://dx.doi.org/10.1021/ac00253a046}
  {\path{doi:10.1021/ac00253a046}}.
\newline\urlprefix\url{https://pubs.acs.org/doi/abs/10.1021/ac00253a046}

\bibitem{galindo-uribarri_high_2021}
A.~Galindo-Uribarri, Y.~Liu, E.~Romero~Romero, D.~W. Stracener,
  \href{https://www.nature.com/articles/s41598-021-01886-z}{High efficiency
  laser resonance ionization of plutonium}, Scientific Reports 11~(1) (2021)
  23432.
\newblock \href {http://dx.doi.org/10.1038/s41598-021-01886-z}
  {\path{doi:10.1038/s41598-021-01886-z}}.
\newline\urlprefix\url{https://www.nature.com/articles/s41598-021-01886-z}

\bibitem{kunz_efficient_2004}
P.~Kunz, G.~Huber, G.~Passler, N.~Trautmann,
  \href{http://link.springer.com/10.1140/epjd/e2004-00024-4}{Efficient
  three-step, two-color ionization of plutonium using a resonance enhanced
  2-photon transition into an autoionizing state}, The European Physical
  Journal D - Atomic, Molecular and Optical Physics 29~(2) (2004) 183--188.
\newblock \href {http://dx.doi.org/10.1140/epjd/e2004-00024-4}
  {\path{doi:10.1140/epjd/e2004-00024-4}}.
\newline\urlprefix\url{http://link.springer.com/10.1140/epjd/e2004-00024-4}

\bibitem{lassen_current_2017}
J.~Lassen, R.~Li, S.~Raeder, X.~Zhao, T.~Dekker, H.~Heggen, P.~Kunz, C.~D.
  P.~Levy, M.~Mostanmand, A.~Teigelhöfer, F.~Ames,
  \href{http://link.springer.com/10.1007/s10751-017-1407-9}{Current
  developments with {TRIUMF}’s titanium-sapphire laser based resonance
  ionization laser ion source: {An} overview}, Hyperfine Interactions 238~(1)
  (2017) 33.
\newblock \href {http://dx.doi.org/10.1007/s10751-017-1407-9}
  {\path{doi:10.1007/s10751-017-1407-9}}.
\newline\urlprefix\url{http://link.springer.com/10.1007/s10751-017-1407-9}

\bibitem{langmuir_thermionic_1925}
I.~Langmuir, K.~H. Kingdon,
  \href{https://royalsocietypublishing.org/doi/10.1098/rspa.1925.0005}{Thermionic
  effects caused by vapours of alkali metals}, Proceedings of the Royal Society
  of London. Series A, Containing Papers of a Mathematical and Physical
  Character 107~(741).
\newblock \href {http://dx.doi.org/10.1098/rspa.1925.0005}
  {\path{doi:10.1098/rspa.1925.0005}}.
\newline\urlprefix\url{https://royalsocietypublishing.org/doi/10.1098/rspa.1925.0005}

\bibitem{dombsky_isac_2003}
M.~Dombsky, P.~Bricault, P.~Schmor, M.~Lane,
  \href{https://linkinghub.elsevier.com/retrieve/pii/S0168583X0201902X}{{ISAC}
  target operation with high proton currents}, Nuclear Instruments and Methods
  in Physics Research Section B: Beam Interactions with Materials and Atoms 204
  (2003) 191--196.
\newblock \href {http://dx.doi.org/10.1016/S0168-583X(02)01902-X}
  {\path{doi:10.1016/S0168-583X(02)01902-X}}.
\newline\urlprefix\url{https://linkinghub.elsevier.com/retrieve/pii/S0168583X0201902X}

\bibitem{friedrich_limits_2021}
S.~Friedrich, G.~B. Kim, C.~Bray, R.~Cantor, J.~Dilling, S.~Fretwell, J.~A.
  Hall, A.~Lennarz, V.~Lordi, P.~Machule, D.~McKeen, X.~Mougeot, F.~Ponce,
  C.~Ruiz, A.~Samanta, W.~K. Warburton, K.~G. Leach,
  \href{https://link.aps.org/doi/10.1103/PhysRevLett.126.021803}{Limits on the
  {Existence} of sub-{MeV} {Sterile} {Neutrinos} from the {Decay} of {Be} in
  {Superconducting} {Quantum} {Sensors}}, Physical Review Letters 126~(2)
  (2021) 021803.
\newblock \href {http://dx.doi.org/10.1103/PhysRevLett.126.021803}
  {\path{doi:10.1103/PhysRevLett.126.021803}}.
\newline\urlprefix\url{https://link.aps.org/doi/10.1103/PhysRevLett.126.021803}

\bibitem{kunz_medical_2020}
P.~Kunz, C.~Andreoiu, V.~Brown, M.~Cervantes, J.~Even, F.~H. Garcia,
  A.~Gottberg, J.~Lassen, V.~Radchenko, C.~F. Ramogida, A.~K.~H. Robertson,
  P.~Schaffer, R.~Sothilingam,
  \href{https://www.epj-conferences.org/10.1051/epjconf/202022906003}{Medical
  isotope collection from {ISAC} targets}, EPJ Web of Conferences 229 (2020)
  06003.
\newblock \href {http://dx.doi.org/10.1051/epjconf/202022906003}
  {\path{doi:10.1051/epjconf/202022906003}}.
\newline\urlprefix\url{https://www.epj-conferences.org/10.1051/epjconf/202022906003}

\bibitem{fiaccabrino2021}
D.~E. Fiaccabrino, P.~Kunz, V.~Radchenko,
  \href{https://www.sciencedirect.com/science/article/pii/S0969805121000184}{Potential
  for production of medical radionuclides with on-line isotope separation at
  the {ISAC} facility at {TRIUMF} and particular discussion of the examples of
  {165Er} and {155Tb}}, Nuclear Medicine and Biology 94-95 (2021) 81--91.
\newblock \href
  {http://dx.doi.org/https://doi.org/10.1016/j.nucmedbio.2021.01.003}
  {\path{doi:https://doi.org/10.1016/j.nucmedbio.2021.01.003}}.
\newline\urlprefix\url{https://www.sciencedirect.com/science/article/pii/S0969805121000184}

\bibitem{ziegler_srim_2010}
J.~F. Ziegler, M.~D. Ziegler, J.~P. Biersack,
  \href{https://linkinghub.elsevier.com/retrieve/pii/S0168583X10001862}{{SRIM}
  – {The} stopping and range of ions in matter (2010)}, Nuclear Instruments
  and Methods in Physics Research Section B: Beam Interactions with Materials
  and Atoms 268~(11-12) (2010) 1818--1823.
\newblock \href {http://dx.doi.org/10.1016/j.nimb.2010.02.091}
  {\path{doi:10.1016/j.nimb.2010.02.091}}.
\newline\urlprefix\url{https://linkinghub.elsevier.com/retrieve/pii/S0168583X10001862}

\bibitem{bortels_analytical_1987}
G.~Bortels, P.~Collaers,
  \href{https://linkinghub.elsevier.com/retrieve/pii/0883288987901808}{Analytical
  function for fitting peaks in alpha-particle spectra from {Si} detectors},
  International Journal of Radiation Applications and Instrumentation. Part A.
  Applied Radiation and Isotopes 38~(10) (1987) 831--837.
\newblock \href {http://dx.doi.org/10.1016/0883-2889(87)90180-8}
  {\path{doi:10.1016/0883-2889(87)90180-8}}.
\newline\urlprefix\url{https://linkinghub.elsevier.com/retrieve/pii/0883288987901808}

\bibitem{james_minuit_1975}
F.~James, M.~Roos,
  \href{https://linkinghub.elsevier.com/retrieve/pii/0010465575900399}{Minuit -
  a system for function minimization and analysis of the parameter errors and
  correlations}, Computer Physics Communications 10~(6) (1975) 343--367.
\newblock \href {http://dx.doi.org/10.1016/0010-4655(75)90039-9}
  {\path{doi:10.1016/0010-4655(75)90039-9}}.
\newline\urlprefix\url{https://linkinghub.elsevier.com/retrieve/pii/0010465575900399}

\bibitem{isacyielddb}
P.~Kunz, \href{https://yield.targets.triumf.ca}{{TRIUMF} {Isotope} {Database}}
  (Jun. 2022).
\newline\urlprefix\url{https://yield.targets.triumf.ca}

\bibitem{rodriguez_2017}
J.~L. Rodr\'{\i}guez-S\'anchez, J.-C. David, D.~Mancusi, A.~Boudard, J.~Cugnon,
  S.~Leray,
  \href{https://link.aps.org/doi/10.1103/PhysRevC.96.054602}{Improvement of
  one-nucleon removal and total reaction cross sections in the {Li\`ege}
  intranuclear-cascade model using {Hartree-Fock-Bogoliubov} calculations},
  Phys. Rev. C 96 (2017) 054602.
\newblock \href {http://dx.doi.org/10.1103/PhysRevC.96.054602}
  {\path{doi:10.1103/PhysRevC.96.054602}}.
\newline\urlprefix\url{https://link.aps.org/doi/10.1103/PhysRevC.96.054602}

\bibitem{kelic_abla07_2008}
A.~Kelic, M.~V. Ricciardi, K.~H. Schmidt,
  \href{http://inis.iaea.org/search/search.aspx?orig_q=RN:40048002}{{ABLA07} -
  {Towards} a complete description of the decay channels of a nuclear system
  from spontaneous fission to multifragmentation}, Tech. rep., International
  Atomic Energy Agency (IAEA), {INDC(NDS)}--0530, pages 181--221 (2008).
\newline\urlprefix\url{http://inis.iaea.org/search/search.aspx?orig_q=RN:40048002}

\end{thebibliography}

\end{document}